\documentclass[11pt,twoside]{article}

\usepackage{asp2004}
\usepackage{epsf}
\usepackage{psfig}
\usepackage{lscape}

\def\be{\begin{displaymath}}
\def\ee{\end{displaymath}}
\def\mas{\mbox{\,M$_\odot$}}

\begin{document}
\title{On the Origin of the most massive stars around R136}
\author{Bernhard R. Brandl}  
\affil{Leiden Observatory, P.O. Box 9513, 2300 RA Leiden, The Netherlands}   
\author{Simon Portegies-Zwart}
\affil{University of Amsterdam}
\author{Anthony F. J. Moffat}
\affil{D\'{e}partment de physique, Universit\'{e} de Montr\'{e}al,  
       Montr\'{e}al, Canada}
\author{David F. Chernoff}
\affil{Astronomy Department, Cornell University, Ithaca, NY 14853} 

% ===============================================================

\begin{abstract} %%% Abstract to run on from here.
We discuss the signature of a peculiar constellation of very massive
stars at a projected distance of 2-3~pc around R136a. We discuss
various scenarios for its possible origin, such as independent
clusters, triggered star formation, supernovae, and ejections via
dynamical interactions.  If the latter scenario were the correct one
this would have significant implications on the way to probe the
conditions in dense stellar cores, and on the evolution of massive
clusters in general.
\end{abstract}

% ===============================================================

\section{Observational findings}
\subsection{Introduction}
The 30\,Doradus region in the LMC is the largest and most massive
H\,II region in the Local Group at a distance of 53~kpc.  Within a
diameter of $15'$ (200~pc) it contains more than $8\times 10^5\mas$ of
ionized gas \cite{ken84}.  The inner 60~pc contain the stellar cluster
NGC\,2070 with about 2400 OB stars, which are responsible for about
1/3 of the total ionizing radiation of the entire 30\,Doradus region
\cite{par93}.  The dense center of NGC\,2070 is called R136
(HD\,38268).  With a bolometric luminosity of $7.8\times 10^7
L_{\odot}$ \cite{mal94} in the inner 4.5~pc R136 is often considered
the closest example of a starburst region.
\begin{figure}[ht]\center
\vbox{\psfig{figure=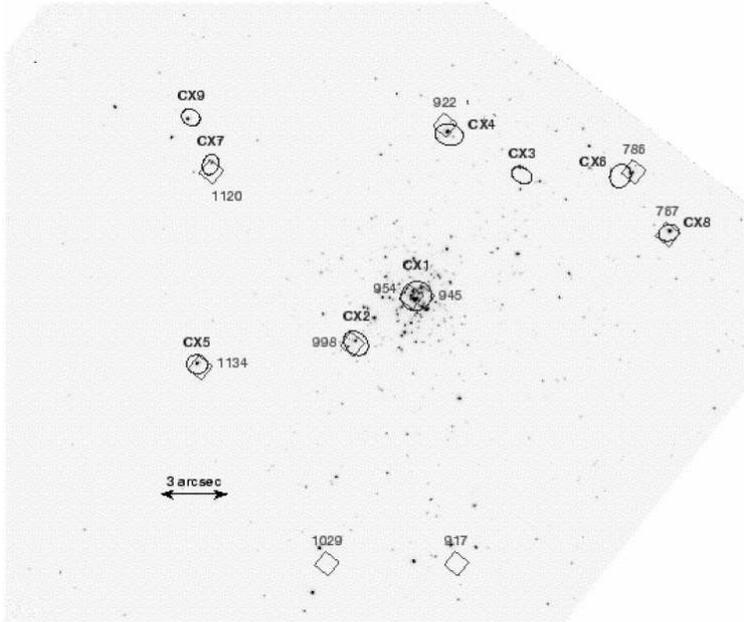,width=10.0cm,angle=-90}
\caption{\small HST image of R136, the central portion of the 30
         Doradus region \cite{mas98}. The exposure time was
         26~s with filter F336W (centered on $3344\AA$ and with a $381\AA$
         bandpass) using WFPC2. The Wolf-Rayet stars identified by
         \cite{par93} are indicated by squares, and the ellipses give
         the locations of the Chandra X-ray sources.  The sizes of the
         ellipses give the $1\sigma$ positional accuracy for the
         Chandra sources \cite{por02}.\label{fig1}}}
\end{figure}

\subsection{Observational peculiarity}
Figure~\ref{fig1} shows the projected distribution of emission
line stars in the central part of NGC\,2070.  For the purpose of this
study their exact spectral classification is rather subordinate; what
is important here, however, is that these stars are very
massive O-stars (see Table~\ref{tab1}).  The stars labeled in
Fig.~\ref{fig1} either have been classified as Wolf-Rayet stars
\cite{par93} or X-ray sources, presumable HMXBs \cite{por02}.

Table~\ref{tab1} lists the projected distance and spectral types for
these stars.  About one third of them is located within $4''$ of
R136a, close to another third is spread across the entire region, with
distances ranging from $16''\ldots 135''$. But more than one third of
these massive stars in 30\,Dor appear to be located in a shallow
sphere of $r\approx 8'' - 13''$ (although even more of the massive
stars, which we attribute to the core region, may be actually located
in the sphere behind or in front of R136a).

\begin{table}[ht]\center
\scriptsize
\begin{tabular}{l l c || l l c }
\\
\multicolumn{3}{c ||}{\bf Stars within and far from R136} & 
\multicolumn{3}{c}{\bf Stars within the ``ring''}\\
{\bf Name} & {\bf Sp. type} & {\bf Distance} & {\bf Name} & {\bf Sp. type} & 
{\bf Distance}\\
Mk30 & O3If*/WN6-A & $16''\le d \le 135''$ & Mk33 & O3If* & $12.8''$ \\ 
Mk51 & O4If & $16''\le d \le 135''$ & Mk33S & O3III(f*) & $11.2''$ \\ 
Mk53a & WN8 & $16''\le d \le 135''$ & Mk34 & WN4.5 & $10.2''$ \\ 
R145 & WN6 & $16''\le d \le 135''$ & Mk35 & O3If*/WN-6A & $11.9''$ \\ 
R140N & WC5+WN4 & $16''\le d \le 135''$ & Mk37W & O4If+ & $11.2''$ \\ 
R140S & WN4.5 & $16''\le d \le 135''$ & Mk37 & O4If+ &  \\ 
R144 & WN6-A(B) & $16''\le d \le 135''$ & Mk39 & O3If*/WN-6A & $12.3''$ \\ 
R146 & WN4+OB & $16''\le d \le 135''$ & Mk42 & O3If*/WN-6A & $8.1''$ \\
R136a1 & WN4.5 & $d \le 4''$ & R134 & WN6 & $11.7''$ \\ 
R136a2 & WN4 & $d \le 4''$ & & & \\ 
R136a3 & WN4.5 & $d \le 4''$ & & & \\ 
R136a5 & O3If*/WN6-A & $d \le 4''$ & & & \\ 
R136a7 & O3III(f*) & $d \le 4''$ & & & \\
R136b & & $d \le 4''$ & & & \\ 
R136c & & $d \le 4''$ & & & \\
\end{tabular}
\caption{Emission line stars in and near R136.  The spectral type is
         based on determinations by \cite{mas98}.\label{tab1}}  
\normalsize
\end{table}

Could this be just a statistical fluctuation?  For a core radius
of $r_c\approx 1''$ \cite{bra96} and under the assumption that
the stellar density distribution follows roughly a standard
cluster King profile, we would expect
\be 
\int_{8''}^{13''}\frac{I_0}{\sqrt{1+(\frac{r}{r_c})^2}}dr = I_0 r_c
\left[ \ln (r+\sqrt{r^2 + r_c^2}) \right]_{8''}^{13''} = 2.2
\mbox{\ stars,}
\ee
in the projected sphere, while we find 9 $(\pm 3)$ emission line stars
-- well above a random statistical fluctuation! Thus we conclude that
their apparent location must have a physical origin.

% ===============================================================
\section{Discussion of various scenarios}
\subsection{The stars were born at their observed location}  

\subsubsection{Independent smaller clusters}
Generally, massive stars do not form in isolation but in clusters and
associations.  For a Salpeter IMF \cite{sal55} one would expect about
65 stars with $m \ge 4\mas$ for each star of $m\ge 60\mas$.  Are the
massive stars around R136 surrounded by increased stellar density,
relative to the overall cluster density profile, or do they appear to
be isolated?

\begin{figure}[ht]\center
\vbox{\psfig{figure=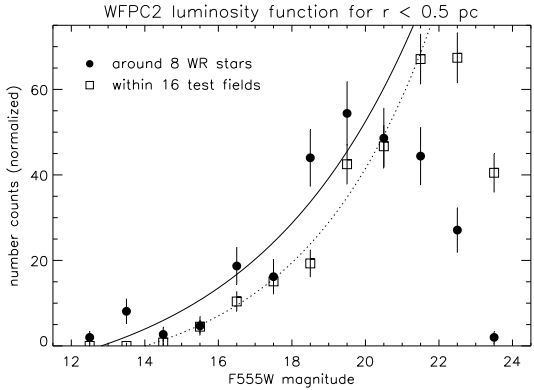,width=7.0cm,angle=0}
\caption{\small Local luminosity functions for 16 randomly chosen test
         fields (8 closer to the core and 8 farther out than the
         ``sphere'') and the emission line stars, within 0.5pc
         ($2''$).  The number counts were normalized to the overall
         cluster density profile and the results from the 16 test
         fields were averaged.  The same approach was followed to
         derive the average luminosity function of stars around the 8
         emission line stars in the sphere.  Beyond $V=20.5$ the data
         become incomplete in the vicinity of a bright source.  The
         data are WFPC2 observations in the $V$ (F555W) and $I$
         (F814W) filter bands \cite{hun96}.\label{fig2}}}
\end{figure}

Figure~\ref{fig2} shows the local luminosity functions, derived from
WFPC2 data in the V (F555W) and I (F814W) filter bands \cite{hun96},
for 16 randomly chosen test fields and the 9 emission line stars.
There is no noticeable enhancement in stellar density around the
peculiar emission line stars that would support the presence of
smaller clusters.  However, we note the possibility that smaller
associations could have dissolved within only a couple of million
years to a level, which is below the detection limit.

\subsubsection{Triggered star formation}
Sequential star formation has been found to have triggered the LMC
associations LH9 and LH10 \cite{par92} as well as N144 and N158
\cite{lor88} in the vicinity of the 30\,Doradus nebula. Several sites
of ongoing star formation, such as Hodge~301 and R143, are located
only 3~arcminutes from R136.  However, the peculiar emission line star
are located very close to R136, in a region of very little
extinction where the gas has been mostly cleared out \cite{bra96}.  No
gaseous filamentary structures, often associated with triggered star
formation are present in the inner parts.  Most importantly, the
emission line stars around R136 are likely of the same age than the
stars within R136.  The core of R136 still appears to be the most
recent site of massive star formation within $30''$, making a second
generation of massive stars in its vicinity rather unlikely.

\subsection{The stars were born within R136 and moved outwards}
\subsubsection{Supernova explosions}
Runaway OB~stars, i.e., apparent field stars with high relative
velocities of $\ge 40$~km/s have been known for a long time --- a
prime example are AE~Auriga and $\mu$~Columba both moving at 100~km/s
away from Orion \cite{bla54}.  Unfortunately, relative velocities are
difficult to determine for the stars around R136.  Although it would
only take them 0.3~Myr to reach the observed location from the cluster
core, even at a runaway velocity as low as 10~km/s the corresponding
proper motion would only be $0.\!''001$ over a 20 year
baseline. Radial velocity studies are in progress, but complicated by
the energetic winds from these stars.

At any rate, some of the emission line stars are O-star binaries,
which cannot be explained by supernova explosions.  Most importantly,
with an age of less than 4~Myr, R136 is unlikely to have yet produced
such a large number of supernovae, and even more so when taking the
kinematical age, i.e., the ``travel time'' from the birthplace to the
observed locations, into account.

\subsubsection{Stellar interactions}
One can estimate that the timescales for global mass
segregation of R136 are about $10^8$~yr \cite{bra96}.  However, the
relaxation time for the high mass ($m_H$) stars alone is significantly
shorter than for the low mass ($m_L$) stars by a factor
$\frac{m_L}{m_H}\left(1+\frac{\sigma^2_H}{\sigma^2_L}\right)^{\frac{3}{2}}$,
where $\sigma_L$ and $\sigma_H$ are the velocity dispersions for the
low and high mass stars, respectively.  Hence, even in the simplified
analytic framework, which assumes point-like, identical particles and
no binaries, the relaxation time for the high mass stars can easily be
$1-2$ orders of magnitude shorter.

\begin{figure}[ht]\center
\vbox{\psfig{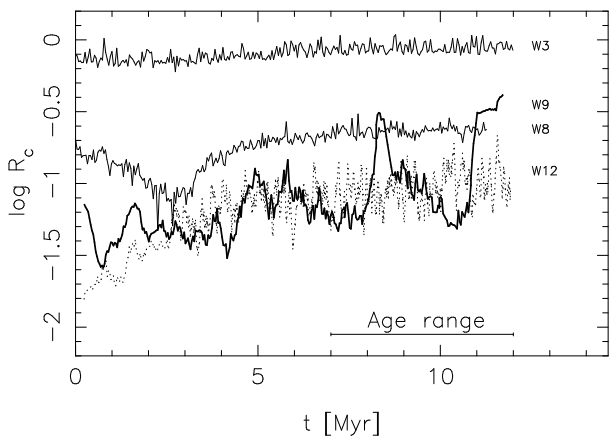}
\caption{\small Evolution of the cluster core radius with time for
         different concentration parameters $(r_c / r_{hm})$.  The
         densest systems (large $W$ numbers) show dramatic
         variations \cite{por04}. \label{fig3}}}
\end{figure}

Recently, Portegies Zwart et al.\ (2004) performed N-body simulations
of a much more realistic situation: 130,000 stars with a mass
distribution given by a Salpeter IMF, a cluster half-mass radius of
1.2~pc, and an initial binary frequency of $30\%$\footnote{The binary
frequency of stars, in particular high mass stars, is a very difficult
problem to attack, both observationally and theoretically.  Studies in
the Orion cluster by \cite{pre99} using speckle interferometry
indicate a large number of $1.5 - 4$ companions per primary OB star --
about $3 - 8$ times higher than the frequency of low-mass
companions.}.  The initial conditions used in the simulations are
similar to the conditions in R136.  The high central density,
existence of binaries, finite stellar sizes and realistic mass
spectrum allow the cluster to undergo sigificant dynamical evolution
early on -- provided that the core density is sufficiently high.
Figure~\ref{fig3} shows the evolution of the cluster core radius for
different concentration parameters.  In the most extreme case, core
collapse happens at less than 1~Myr.  At those high stellar densities
dynamical processes (binary-binary, or binary-single star
interactions) can account for numerous ejections of massive stars from
R136.
%

% ===============================================================
\section{Open questions} 

\subsubsection{Can massive binaries be ejected and become bright X-ray 
               sources?}  
Whether an ejected binary can become a luminous X-ray source depends
on its orbital separation.  For a total mass of R136 of $5\cdot
10^5\mas$, a mean stellar mass of $0.5\mas$, and a half-mass radius of
1.2~pc a binary of mass $m_b = 40\mas$ and a binding energy $E_{bind} =
2000 k T$ will have an orbital separation of $d = 222 R_\odot$, which
is a typical separation for an X-ray binary.

\subsubsection{Why is this signature not observed in NGC\,3603?}
Although NGC\,3603 has been referred to a Galactic clone of R136
\cite{mof94} there are significant differences.  Most relevant here is
that NGC\,3603, with an age of about 1~Myr \cite{bra99}, is
significantly younger than R136.  Given the strong dependency on
central density and quick evolution (see Fig.\ref{fig3}) a few hundred
thousand years can make a big difference.
 
\subsubsection{Why should all ejected stars be observed at the same distance?}
First, the tidal radius of R136 is about $21''$ \cite{mey93}.  {\em
If} the stars are still bound to the cluster potential they are in
elliptical orbits around R136 and most likely to be found at the
apocenter.  Second, most of the dynamical interactions, which have led
to an ejection of a massive star, have occurred within a narrow time
window, namely after the core has reached a critical density.  Third,
even spherically uniform distributions tend to appear ring-like in
projection.

% ================================================================
% ----------------------BIBLIOGRAPHY------------------------

\end{document}